Quantification of $CO_2$ generation in sedimentary basins through Carbonate/Clays Reactions with uncertain thermodynamic parameters

by, G. Ceriotti[a], G.M. Porta[a], C. Geloni[b] and M. Dalla Rosa[b], A. Guadagnini[a,c]

[a]Department of Civil and Environmental Engineering, Politecnico di Milano, Piazza L. Da Vinci 32, 20133 Milano, Italy

[b]Eni S.p.A.-Upstream and Technical Services, via Emilia, 1 20097 San Donato Milanese (MI) Italy

[c] Department of Hydrology and Atmospheric Sciences, University of Arizona, Tucson, AZ 85721, USA

Corresponding author: Giulia Ceriotti, Department of Civil and Environmental Engineering, Politecnico di Milano, Piazza L. Da Vinci 32, 20133 Milano, Italy, giulia.ceriotti@polimi.it, phone 0039 2399 6257.

# Abstract

We develop a methodological framework and mathematical formulation which yields estimates of the uncertainty associated with the amounts of $CO_2$ generated by carbonate-clays reactions (CCR) in large-scale subsurface systems to assist characterization of the main features of this geochemical process. Our approach couples a one-dimensional compaction model, providing the dynamics of the evolution of porosity, temperature and pressure along the vertical direction, with a chemical model able to quantify the partial pressure of $CO_2$ resulting from minerals and pore water interaction. The modeling framework we propose allows (*i*) estimating the depth at which the source of gases is located and (*ii*) quantifying the amount of $CO_2$ generated, based on the mineralogy of the sediments involved in the basin formation process. A distinctive objective of the study is the quantification of the way the uncertainty affecting chemical equilibrium constants propagates to model outputs, i.e., the flux of $CO_2$. These parameters are considered as key sources of uncertainty in our modeling approach because temperature and pressure distributions associated with deep burial depths typically fall outside the range of validity of commonly employed geochemical databases and typically used geochemical software. We also analyze the impact of the relative abundancy of primary phases in the sediments on the activation of CCR processes. As a test bed, we consider a computational study where pressure and temperature conditions are representative of those observed in real sedimentary formation. Our results are conducive to the probabilistic assessment of (*i*) the characteristic pressure and temperature at which CCR leads to generation of $CO_2$ in sedimentary systems, (*ii*) the order of magnitude of the $CO_2$ generation rate that can be associated with CCR processes.

**Introduction**

Natural accumulations of $CO_2$ are commonly observed in sedimentary basins. The carbon dioxide occurs as a gaseous phase with proportions ranging from 5% to 100% of the total gas phase volume. These $CO_2$ accumulations are exploited in several sectors, including, for example, the food industry (e.g., Broadhead et al., 2009) or within the context of Enhanced Oil Recovery (EOR; e.g., Allis et al., 2001 and references therein) operations. They are also investigated as natural analogs for improving our understanding and design of subsurface $CO_2$ storage protocols/technologies and for the assessment of the ensuing environmental risks associated with diverse migration pathways connecting sources with receptors (e.g., Metz et al., 2005). Accumulation of $CO_2$ in sedimentary basins can lead to dilution of valuable hydrocarbon gas mixtures (e.g., methane and propane), thus reducing energy storage in a reservoir and resulting in an increased production cost (Imbus et al., 1998).

Various authors indicate diverse organic and/or inorganic processes as possible causes of natural $CO_2$ accumulation (e.g., Higgs et al., 2013; Hutcheon and Abercrombie, 1990; Clayton et al., 1990; van Berk et al., 2013; Smith and Ehrenberg, 1989; Chiodini et al., 2007; Ballentine et al., 2001; Cooper et al. 1997; Dubacq et al., 2012; Fischer et al. 2006; Houtcheon et al., 1980; Imbus et al, 1998; Cathles and Schoell, 2007; Kotarba and Nagao, 2008; Li et al, 2008; Mayo and Muller. 1997; Wycherley et al. 1999; Cai et al. 2001; Farmer, 1965; Goldsmith 1980 Arnórsson, 1986, Chiodini et al., 2000; Fischer et al., 2006 ). Among these sets of processes, in this study we focus on $CO_2$ generation in sedimentary formations through the Carbonate/Clay Reaction (CCR) mechanism. The role of CCR as a possible relevant $CO_2$ generating mechanism in sedimentary systems is originally suggested by Hutcheon and Abercrombie (1990), Hutcheon et al. (1980), Hutcheon et al. (1990), Hutcheon et al. (1993). The feasibility of CCR occurrence in a sedimentary environment is supported by a significant amount of studies (e.g., Coudrain-Ribstein and Gouze, 1993, Coudrain-Ribstein et al., 1998, Cathles and Schoell, 2007; Giggenbach, 1980; Smith and Ehrenberg, 1989; Chiodini et al., 2007; Xu and Pruess, 2001; van Berk et al., 2009). These works document a series of field data about $CO_2$ partial pressure and/or pore-water chemical compositions sampled in real sedimentary basins and/or computed through geochemical speciation models which are compatible with the CCR mechanism. Cathles and Schoell (2007) propose a clear and schematic conceptual model of $CO_2$ generation through CCR and provide a mathematical formulation relying on a chemical equilibrium model for the identification of the environmental conditions (temperature and pressure) at which $CO_2$ may be generated as a separate gas phase. These authors illustrate the use of their model through an exemplary setting assuming a time-invariant linear relationship between temperature and pressure, along the lines of Smith and Ehrenberg (1989). The results of this study suggest that CCR may become a relevant process for gaseous $CO_2$ generation at a temperature of about 330 °C. Even as the results of the illustrative example of Cathles and Schoell (2007) are not directly transferable to a real sedimentary basin setting (where temperature and pressure vary with time according to a higher complexity pattern), they clearly suggests that the $CO_2$ gas generation associated with CCR is expected to occur at very high temperatures and pressures.

Uncertainties associated with thermodynamic parameters characterizing CCR are virtually ubiquitous. This is the consequence of a variety of factors (including, e.g., intrinsic natural variability of mineral compositions, non-ideal behavior of multiphase solutions, paucity and/or inaccuracy of available experimental data) and constitutes a critical challenge for the robust characterization of geochemical processes taking place at high temperature and pressure which are typically observed in deep sedimentary formations.

In this context, the major objective of our study is to propose a general framework within which we develop a modeling approach which incorporates the uncertainty associated with the thermodynamic parameters characterizing the CCR mechanism to yield a quantitative estimation of the amount of $CO_2$ released from CCR in sedimentary formations. Our approach is grounded on two coupled components: (*i*) a compaction model, simulating the burial history of a sedimentary basin; and (*ii*) a geochemical model which quantifies the amount of generated $CO_2$ (as a dissolved or separate gaseous/supercritical phase) on the basis of thermodynamic equilibrium concepts. For the purpose of demonstrating our approach, we consider the one-dimensional compaction model presented by Formaggia et al. (2013), Porta et al. (2014) and Colombo et al. (2016), other numerical models (eventually characterized by an increased degree of complexity) being fully compatible with our methodological framework. Quantification of $CO_2$ in aqueous and gaseous phase in surface environments or shallow subsurface systems is generally tackled through a hydro-geochemical speciation software (e.g., Phreeqc, Parkhurst and Appelo, 2013). Available databases supporting these software are typically considered as reliable within a range of temperatures lower than 300 °C. Settings of the kind investigated in this study are characterized by temperatures larger than 300 °C and pressure values significantly larger than those typically found in shallow aquifer systems. Hence, we employ here an *ad-hoc* geochemical model which is consistent with the formulations proposed by Giggenbach (1981), Coudrain-Ribstein et al. (1998) and Cathles and Schoell (2007) and can be applied in the presence of temperature/pressure conditions taking place in deep sedimentary formations.

We highlight that a major element of novelty of our work is the analysis of the way uncertainties associated with the thermodynamic parameters employed to characterize the CCR mechanism, i.e., the mineral solubility and phase equilibrium constants, propagate to the final model outputs. These parameters are viewed as random model inputs characterized by a given probability density function (*pdf*). As a consequence, all outputs are considered in a probabilistic framework. A variety of additional sources of model and parametric uncertainty (Neuman, 2003) may affect the outputs of the proposed modeling approach. These include, e.g., the salinity of the brine and the feedback with other geochemical processes which may take place in sedimentary systems. In this work we focus on the characterization of parametric uncertainty related to thermodynamic equilibrium constants, because these parameters are not firmly constrained at the pressure and temperature conditions of interest. To the best of our knowledge, an assessment of this kind is still lacking in the context of basin scale modeling of CCR processes.

Key target quantities that we consider as model outputs are the amount of $CO_2$ produced in the system and its temporal dynamics resulting from the compaction processes of the sediment evolving along geologic time scales. Results stemming from our approach include an explicit quantification of the depth at which the source of gaseous $CO_2$ is located and of the impact of the relative abundance of primary phases affecting the generation of $CO_2$. As a first test bed to illustrate our methodology, we implement the conceptual and numerical model proposed on a realistic sedimentary basin setting in terms of temperature-pressure-porosity, upon considering multiple scenarios in terms of relative abundance of CCR primary phases in the mineralogical assemblage. We base this study on a streamlined conceptual and numerical model of the system to allow (*i*) focusing on the stochastic analysis of selected uncertain quantities and (*ii*) comparing our results against available literature data. As such, we consider uncertainty to be embedded in the effects of the temperature on the thermodynamic constants regulating the equilibrium between $CO_2$ - water - mineral phases, all of the remaining model features being treated as deterministic (see also Section 3 for a detailed discussion). The methodological framework we propose is then portable to scenarios characterized by an increased level of complexity and in the presence of a variety of sources of uncertainty.

The work is structured as follows: in Section 1 we provide a brief overview of the CCR process; Section 2 illustrates the theoretical framework and modeling workflow as well as the coupled formulation of the geochemical and basin models we employ to quantify the $CO_2$ generated during the basin evolution; in Section 3 we illustrate the main sources of uncertainty which can arise in our modeling procedure and classify these into modeling and parametric uncertainties; in Section 4 we present the main results obtained by the implementation of the modeling workflow for a basin-scale case study; Section 5 is devoted to a detailed discussion and analysis of the results. We provide conclusions and an overview on future perspectives in Section 6.

# 1   Overview of CCR processes

Previous works (e.g. Giggenbach , 1980) have shown that the presence of carbonate phases along with clays and/or alumino-silicates in high-temperature geothermal or sedimentary systems acts as a buffer system for the pore-water and might then control the partial pressure of $CO_2$. Assuming that the rock-fluid system attains an equilibrium, we can model the interaction between carbonates, alumino-silicates, clays and $CO_2$ as a single equilibrium reaction. The latter is typically termed Carbonate/Clays Reaction (CCR), following the nomenclature introduced by Hutcheon et al. (1980).

Several authors (Giggenbach, 1978; Giggenbach, 1981; Giggenbach, 1984, Coudrain-Ribstein et al., 1998; Cathles and Schoell, 2007; Hutcheon and Abercrombie, 1990; Hutcheon et al., 1980; Hutcheon at al., 1989; Hutcheon et al., 1993; Zhang et al., 2000; Huang and Longo, 1994; Ueda et al., 2005) suggest a variety of chemical equilibrium relationships to depict the stoichiometry of CCR. Table 1 lists a set of CCRs following the study of Coudrain-Ribstein et al. (1998). These can be generalized through a chemical equilibrium relationship of the kind

$$\alpha_1 M_1 + ... + \alpha_m M_m = \alpha_{m+1} M_{m+1} + ... + \alpha_{n+m} M_{n+m} + \alpha_0 CO_{2(g)} \qquad (1)$$

where $M_k$ ($k = 1, ..., n+m$) represents the $k^{th}$ mineral phase involved in the process and $\alpha_k$ are stoichiometric coefficients.

Since the generation of $CO_2$ is the result of the consumption of the mineral phases of the left hand side of (1), we indicate $M_k$ with $1 \leq k \leq m$ or $m + 1 \leq k \leq m + n$ as primary or secondary phases of the system, respectively. Primary phases include a carbonate mineral (which represents the source of $CO_2$) together with other cations (e.g., $Mg^{+2}$ and $Ca^{+2}$), and additional clay/alumino-silicate phases. The latter act as source of other ions (e.g., $OH^-$, $Al^{3+}$ and $K^+$) when dissolved in water. All reactions listed in Table 1 include dolomite as carbonate mineral. The secondary phases include $CO_2$, clay minerals (e.g., chlorite, phoglopite, illite; Bergaya and Lagaly, 2013) and other species (e.g., calcite) which act as sinks for the ions released by the primary phases and represent a more stable mineralogical assemblage at large temperature (e.g., $T > 300$ °C), as compared to primary phases (Giggenbach, 1981; Hutcheon and Abercrombie, 1990; Smith and Ehrenberg, 1989). Partitioning among primary and secondary phases in the system can be described through the equilibrium constant ($K_R$). All phases appearing in (1) are in pure liquid or solid phases, $CO_2$ being the only gaseous phase. The logarithmic transform of $K_R$ is

$$\log K_R = \alpha_0 \log \eta_{CO2(g)} + \sum_{k=m+1}^{n+m} \alpha_k \log a_{Mk} - \sum_{k=1}^{m} \alpha_k \log a_{Mk} \qquad (2)$$

$a_{Mk}$ and $\eta_{CO2}$ respectively representing the activity of species $M_k$ and the $CO_2$ fugacity. Assuming that the fugacity coefficient of $CO_2$ is equal to one (Hutcheon, 1990, Chiodini et al., 2007; Cathles and Schoell, 2007; Coudrain-Ribstein et al., 1998) yields

$$\log P_{CO2} = \log \eta_{CO2(g)} = \frac{\log K_R}{\alpha_0} \qquad (3)$$

$P_{CO2}$ being the partial pressure of $CO_2$ either in the gaseous or supercritical phase. Note that, according to our assumption, the numerical values of $CO_2$ fugacity and partial pressure coincide (Anderson, 2009). We provide additional discussion about the assumption of $\eta_{CO2} = P_{CO2}$ in Electronic Annex II. The value of $\log K_R$ (and therefore $P_{CO2}$) is a function of the local conditions of pressure and temperature, as discussed in Sections 2.2.

The CCR process can be summarized by the phenomenological scheme illustrated in Fig. 1 and described in the following.

1. Given a sedimentary rock containing carbonates and clays/alumino-silicates, the amount of dissolved $CO_2$ in the pore water is regulated by the chemical equilibrium among all phases (Fig. 1a). Even as a separate gas phase is not formed, the concept of partial pressure associated with gaseous species can be still preserved if referred to a fictive gas phase hypothetically at equilibrium with the pore water (Coudrain-Ribstein et al., 1998).

2. Pressure and temperature typically increase throughout the burial process. Under these conditions, the sum of the partial pressures associated with gaseous species ($CO_2$ and possibly other species including, e.g., $H_2O_{(g)}$, $CH_{4(g)}$) might exceed the fluid environmental pressure. When this happens, a separate gas phase is generated (Fig.1b). In this work we consider only $CO_2$ and $H_2O$ as possible gaseous species.

3. When $CO_2$ (possibly mixed with other gases) is released as a gas phase, the difference between gas and fluid phase densities promotes upward migration of $CO_2$. As a consequence of this migration, the equilibrium reaction (1) is shifted towards its right side (Fig. 1c) and the reactions listed in Table 1 can be considered as a quantitative transformation of the reactants (primary phases) into the products (secondary phases), as seen, e.g., in Cathles and Schoell (2007).

Note that supercritical $CO_2$ is likely to be expected at the pressure and temperature conditions characterizing sedimentary formations. We assume that the conceptual model proposed by Cathles and Schoell (2007) still holds when $CO_2$ is in supercritical conditions. Supercritical $CO_2$ is always characterized by lower density when compared to water and buoyancy effects always force the $CO_2$-rich separate gas phase to migrate upwards as soon as it is generated (Battistelli et al., 2016; Span and Wagner, 1996; Johnson et al., 1992). For simplicity, we refer in the following to the separate $CO_2$-rich phase as gaseous $CO_2$.

When the conditions for the generation of a separate gas phase are not attained, CCR leads only to the formation of aqueous $CO_2$ which remains dissolved in the pore-fluid. Dissolved $CO_2$ can constitute a significant fraction of the overall $CO_2$ amount released by CCR and its occurrence can be a relevant aspect to consider when the characterization of flow processes in sedimentary formations is of concern (Coudrain-Ribstein and Gouze, 1993; Chiodini et al., 2000; Farmer, 1965).

## 2 Modeling workflow

We illustrate here a procedure to compute the time, depths and temperature at which the process described in Fig. 1, i.e. the activation of the gaseous $CO_2$ source, takes place. The two main constituents of the numerical modeling procedure we propose are:

1. a basin compaction model, providing the temporal dynamics of porosity, temperature, pressure and basin stratigraphy along the vertical direction in the presence of mechanical compaction;

2. a geochemical model which allows computing the partial pressure of $CO_2$ ($P_{CO2}$ [Pa]) and the concentration of dissolved $CO_2$ ($C_{CO2}$ [mol/L]) as a function of temperature and pressure.

Our modeling strategy focuses on the uncertainty associated with the identification of $CO_2$ sources and with the quantification of the resulting $CO_2$ fluxes. Characterization of migration of $CO_2$ after its generation is beyond the scope of our study. Fig. 2 illustrates the key steps of the workflow, which is subdivided in three blocks: *i*) implementation of the burial model (Block 1), described in Section 2.1; *ii*) computation of the $CO_2$ pore-water concentration and $CO_2$ partial pressure (Block 2), illustrated in Section 2.2; and *iii*) estimation of $CO_2$ generation rate and source location (Block 3), detailed in Section 2.3. All details on the computational steps of the model are reported in the Electronic Annex I where a step-by-step illustration of the procedure is included to assist reproducibility of the model implementation.

## 2.1 Basin Model

The quantification of the amount of $CO_2$ generated in sedimentary systems requires the quantification of (*i*) porosity ($\phi$), temperature ($T$), pressure ($P$) distributions and burial velocity of sediments ($V_{SED}$, i.e., rate at which the sediments are displaced along the vertical direction) as a function of depth and time; and of (*ii*) the temporal evolution of the stratigraphy. In this study, we obtain these quantities through the one-dimensional compaction model proposed and tested by Formaggia et al. (2013), Porta et al. (2014), and Colombo et al. (2016). Further details related to the burial model implemented in this work can be found in Electronic Annex I. We highlight here that any type of compaction/diagenesis model (e.g., a three-dimensional model) is compatible with the proposed procedure, provided it renders a characterization of the dynamics of temperature, pressure, porosity and sediment burial velocity in the system. In this study, we consider the outputs of the basin model (e.g., temperature and pressure distributions) as deterministic quantities, consistent with our focus on the quantification of the parametric uncertainty related to the geochemical model. Possible sources of uncertainty affecting the burial model are explicitly discussed in Section 3.

### 2.1.1 Basin scale case study and compaction setting

We illustrate the applicability of our methodological framework for the quantification of the uncertainty associated with estimates of $CO_2$ generation at basin scale by focusing on an exemplary system inspired to a realistic compaction setting.

We consider a basin deposition over a period of 135 Ma (Millions of years before present), from time $t$ = 135 Ma, to present day (i.e., $t$ = 0 Ma). According to our simplified compaction model, we assume the basin to be described as a one-dimensional system along the vertical direction. The paleo-bathimetry is constant and set equal to an elevation of 106 m (the $Z$-axis is considered to point downwards and the sea level to correspond to $Z$ = 0). Carbonate sediments are deposited within the interval ranging from time $t$ = 135 Ma and $t$ = 23 Ma, leading to the formation of carbonate rock layers. Shale and sandy shale sediments are deposited within the period ranging from $t$ = 23 Ma and $t$ = 0 Ma, leading to mudrock after compaction. Sediment deposition rate ($V_D$) at the basin top is imposed as boundary condition. We assume that it varies in time and can be described by a piecewise constant function of time across six time intervals as indicated in Table 2. A given temperature gradient of 32 °C /km is prescribed at the basement. Each sediment type is characterized by a given thermal conductivity of the solid matrix ($K_T$), initial porosity (i.e., porosity at sediment deposition time, $\phi_0$), and vertical compressibility coefficients ($\beta$). We set the parameters $K_T$, $\phi_0$ and $\beta$ to the values listed in Table 3.

We analyze the two possible scenarios of mineral composition associated with carbonate rock listed in Table 4. We highlight that: (*i*) Scenario $S_{dol}$ considers dolomite as the only carbonate mineral present in the rock; (*ii*) Scenario S$_{cal}$ is characterized by the presence of magnesiac limestone where calcite is the prevailing carbonate (73% in weight) and the fraction of dolomite is lower than 10.

## 2.2 Geochemical modeling under uncertainty

The main physical quantities which allow quantifying the *CO$_2$* generated by CCR are the partial pressure of the gas phases and the concentration of the *CO$_2$* dissolved in the fluid phase. We obtain these outputs starting from (1)-(3) and relying on the assumption that the activity of the pure solid mineral and liquid phases are set equal to unity (Giggenbach, 1980; Giggenbach, 1984).

The dependence of *P$_{CO2}$* on temperature in (3) is assessed by relying on the thermochemical parameters collected in a thermodynamic database. Among the databases available in the literature (e.g., LLNL, Delany and Lundeen, 1990; Vminteq, Peterson, 1987; SOLMINEQ, Kharaka et al., 1988), we select the Thermoddem database (Blanc et al., 2012) due to its completeness, traceability of data, and proven internal thermodynamic consistence, especially for the aluminum silicate phases (Blanc et al., 2015).

In the remainder of the work, an uncertain (i.e., random) quantity $\zeta$ is identified with the notation $\tilde{\zeta}$. Our operational procedure relies on the following steps:

1. A set of basis species is selected coherently to the chosen thermodynamic database (e.g., 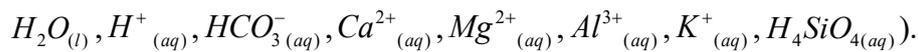
   $H_2O_{(l)}, H^+_{(aq)}, HCO_3^-{}_{(aq)}, Ca^{2+}_{(aq)}, Mg^{2+}_{(aq)}, Al^{3+}_{(aq)}, K^+_{(aq)}, H_4SiO_{4(aq)}$).

2. A set of stoichiometric coefficients are defined to honor mass and charge balances, i.e.,

$$\begin{cases} \gamma_k = \alpha_k & if \quad 1 \leq k \leq m \\ \gamma_k = -\alpha_k & if \quad m+1 \leq k \leq m+n \end{cases} \quad (4)$$

3. A speciation reaction *S$_k$* is defined for each *M$_k$* phase (*k*= 1, …, *n+m*) involved in reaction (1) to describe speciation in the formation fluid (which we consider as water) of *M$_k$* through the basis species selected in step 1. Uncertain chemical equilibrium constants $\tilde{K}_{Sk}$ are quantified to characterize the speciation reaction *S$_k$* of phase *M$_k$* at temperature *T*. We do so by employing the following equation derived from the Maier-Kelley heat capacity definition (Maier and Kelley, 1932; Parkhurst and Appelo, 2013; van Berk et al., 2009)

$$\log \tilde{K}_{Sk} = \tilde{A}_k + \tilde{B}_k T + \frac{\tilde{C}_k}{T} + \tilde{D}_k \log T + \frac{\tilde{E}_k}{T^2} \tag{5}$$

where $\left( \tilde{A}_k, \tilde{B}_k, \tilde{C}_k, \tilde{D}_k, \tilde{E}_k \right)$ is a vector of uncertain quantities which are treated as independent random variables/parameters with an assigned probability density function (*pdf*). The characterization of the uncertainty of these parameters is discussed in Section 2.2.1. Note that the (5) allows evaluating the equilibrium constant $\tilde{K}_{Sk}$ as a function of temperature while keeping the pressure of the system at a constant reference value of 1 bar.

4. The equilibrium constant $\log \tilde{K}_{R,T,1}$ associated with reaction (1) is computed as (Coudrain-Ribstein et al., 1998)

$$\log \tilde{K}_{R,T,1} = \sum_{k=1}^{n} \gamma_k \log \tilde{K}_{Sk} - \alpha_0 \log \tilde{K}_{S,CO2(g)} \tag{6}$$

where $\tilde{K}_{S,CO2(g)}$ is the equilibrium constant associated with the reaction defining the $CO_2$ (either gaseous or supercritical) in terms of its basis species. Note that $\tilde{K}_{S,CO2(g)}$ is considered to be characterized by a relationship having the same format as (5) and is also considered as random. Subscript 1 appearing in $\log \tilde{K}_{R,T,1}$ indicates that the value of the equilibrium constant evaluated through (6) is associated with the reference pressure of 1 bar.

5. We compute $\tilde{K}_{R,T,P}$ as a modification of $\tilde{K}_{R,T,1}$ to account for the high pressure at which the CCR process occurs according to the procedure proposed by Millero (1982). Further details can be found in Electronic Annex I.

6. $CO_2$ partial pressure $\tilde{P}_{CO2}$ is evaluated upon replacing $\log K_R$ with $\log \tilde{K}_{R,T,P}$ in (3).

The activity $\tilde{a}_{CO2(aq)}$ of carbon dioxide dissolved in the liquid phase can be estimate by considering the equilibrium as an effective model

$$CO_{2(g)} = CO_{2(aq)} \qquad (7).$$

Assuming a unit coefficient activity associated with [$CO_{2(aq)}$] (see Electronic Annex II for additional details about this assumption), we can directly derive the molar concentration of aqueous $CO_2$ ($\tilde{C}_{CO2(aq)}$ [mol/l]) from the $CO_2$ activity. A description of the detailed steps leading to the quantification of aqueous $CO_2$ through our computational procedure are included in Electronic Annex I.

Values of the uncertain quantities $\tilde{C}_{CO2(aq)}$ and $\tilde{P}_{CO2}$ may be constrained by the effect of limiting reactants, as a consequence of relative abundancy of diverse primary phases. Given a mineral composition, the generation of $CO_2$ takes place according to the equilibrium relationship (3) until one of the involved primary mineral phases vanishes. We then verify that the computed $\tilde{C}_{CO2(aq)}$ is compatible with the maximum $CO_2$ concentration ($C_{max}$) associated with the complete depletion of the limiting reactant across all primary phases. We set $C_{CO2(aq)} = C_{max}$ at locations where $C_{CO2(aq)}$ is larger than $C_{max}$, and accordingly correct the associated value of $P_{CO2}$. In the following we denote as $\tilde{C}_{CO2(aq)}|C_{max}$ and $\tilde{P}_{CO2}|C_{max}$ the values of dissolved $CO_2$ and $CO_2$ partial pressure conditional to the effect of limiting reactant. Additional computational details related to $C_{max}$, $\tilde{C}_{CO2(aq)}|C_{max}$, and $\tilde{P}_{CO2}|C_{max}$ are included in Electronic Annex I.

### 2.2.1 Characterization of uncertain model inputs

Here we illustrate the stochastic characterization of the chemical equilibrium constants. We assess the consistency of the results stemming from the proposed procedure with available data of $CO_2$ partial pressure in Sections 4.2 and 5.1.

For the purpose of illustration of our uncertainty quantification procedure, hereinafter we focus on the following CCR (see Table 1)

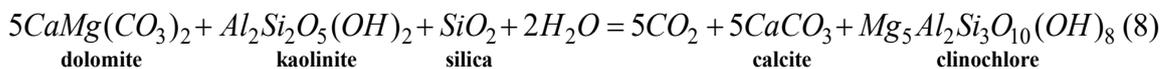

$$\underset{\textbf{dolomite}}{5CaMg(CO_3)_2} + \underset{\textbf{kaolinite}}{Al_2Si_2O_5(OH)_2} + \underset{\textbf{silica}}{SiO_2} + 2H_2O = 5CO_2 + \underset{\textbf{calcite}}{5CaCO_3} + \underset{\textbf{clinochlore}}{Mg_5Al_2Si_3O_{10}(OH)_8} \quad (8)$$

We select this equilibrium reaction among those listed in Table 1 because all of the involved mineral phases are commonly found in sedimentary system (e.g., Hutcheon, 1990; Coudrain-Ribstein and Gouze, 1993) and it is in agreement with the mineralogical assemblage alteration observed in the Kootenay Formation studied by Hutcheon et al. (1980). We follow the procedure outlined in Section 2.2 and start by selecting the following basis species

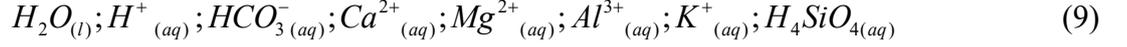

$$H_2O_{(l)}; H^+_{(aq)}; HCO_3^-_{(aq)}; Ca^{2+}_{(aq)}; Mg^{2+}_{(aq)}; Al^{3+}_{(aq)}; K^+_{(aq)}; H_4SiO_{4(aq)} \qquad (9)$$

We then write the following system governing speciation of all liquid and solid phases involved and of the gaseous $CO_2$

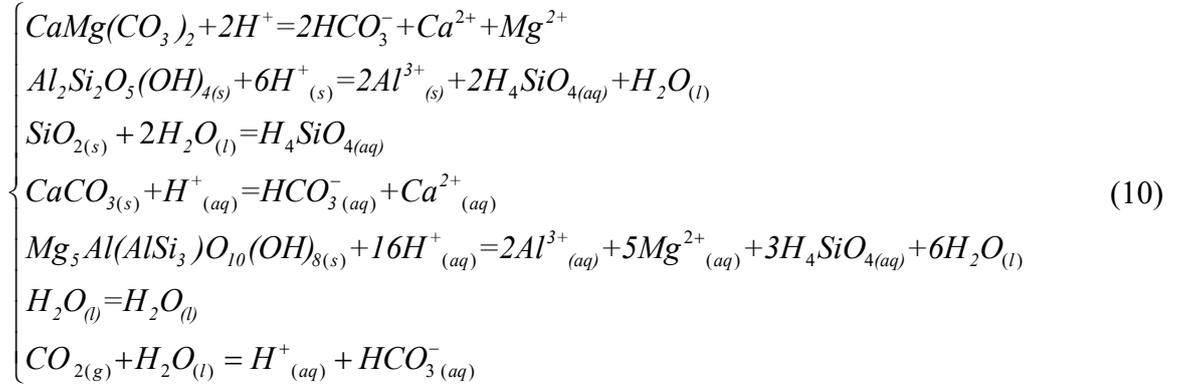

$$\begin{cases} CaMg(CO_3)_2 + 2H^+ = 2HCO_3^- + Ca^{2+} + Mg^{2+} \\ Al_2Si_2O_5(OH)_{4(s)} + 6H^+_{(s)} = 2Al^{3+}_{(s)} + 2H_4SiO_{4(aq)} + H_2O_{(l)} \\ SiO_{2(s)} + 2H_2O_{(l)} = H_4SiO_{4(aq)} \\ CaCO_{3(s)} + H^+_{(aq)} = HCO_3^-_{(aq)} + Ca^{2+}_{(aq)} \\ Mg_5Al(AlSi_3)O_{10}(OH)_{8(s)} + 16H^+_{(aq)} = 2Al^{3+}_{(aq)} + 5Mg^{2+}_{(aq)} + 3H_4SiO_{4(aq)} + 6H_2O_{(l)} \\ H_2O_{(l)} = H_2O_{(l)} \\ CO_{2(g)} + H_2O_{(l)} = H^+_{(aq)} + HCO_3^-_{(aq)} \end{cases} \qquad (10)$$

We characterize the equilibrium constants associated with (10) by relying on Thermoddem as a reference database (Blanc et al., 2012). We employ (5) to account for the influence of temperature, where parameters $(\tilde{A}_k, \tilde{B}_k, \tilde{C}_k, \tilde{D}_k, \tilde{E}_k)$ are assumed to be uncertain.

To streamline the uncertainty quantification procedure, we perform a preliminary sensitivity analysis by means of a numerical Monte Carlo procedure. This enables us to single out the contributions of the five parameters appearing in (5) to the variability of $\tilde{K}_{Sk}$ characterizing the reactions $S_k$ presented in (10). In this framework, the five parameters $(\tilde{A}_k, \tilde{B}_k, \tilde{C}_k, \tilde{D}_k, \tilde{E}_k)$ associated with the quantification of each $\log \tilde{K}_{Sk}$ are randomly sampled from uniform distributions centered on the reference value reported in Thermoddem and of width equal to ±20 % of such reference value. We compute the first order sensitivity index ($SI_h$, $h$ = $A, B, C, D, E$) for each parameter, according to the variance-based method described, e.g., in Sobol (2001), Saltelli et al. (2008), and Razavi and Gupta (2015). These sensitivity indices represent the relative contribution of each uncertain parameter in (5) to the variance of $\log \tilde{K}_{Sk}$. These metrics are widely employed in diverse fields (e.g., Saltelli et al., 2008; Formaggia et al., 2013; Porta et al., 2014; Riva et al., 2015) to perform global sensitivity analysis of the output of a model as driven by a set of uncertain (random) model parameters. The results of this analysis reveal that the variability of $\log \tilde{K}_{Sk}$ computed through (8) is mainly due to the random variability of parameters $\tilde{A}_k$ and $\tilde{D}_k$ (details not shown). These parameters are always associated with first order sensitivity indices ($SI_A$ and $SI_D$) larger than 0.43 for 0 °C ≤ T ≤ 400 °C, $SI_B$, $SI_C$ and $SI_E$ always being lower than 0.02. On these bases, we set parameters $B$, $C$ and $E$ to the values listed in Thermoddem and consider $\tilde{A}_k$ and $\tilde{D}_k$ as random input parameters in (5).

We rely on laboratory scale mineral solubility experiments to characterize the uncertainty associated with $\tilde{A}_k$ and $\tilde{D}_k$. As observed by Blanc et al. (2013), only a few experimental studies reporting values of $\tilde{K}_{Sk}$ as a function of temperature are available, particularly with reference to clay minerals. The estimation of the coefficients embedded in Thermoddem is mostly based on thermodynamic calculations, which are associated with an uncertainty level which is difficult to quantify. Here, we employ the mineral solubility data for calcite and kaolinite reported by Plummer and Busenberg (1982) and Blanc et al. (2013). Plummer and Busenberg (1982) provide observations of the calcite speciation constant for a series of temperatures ranging from 0.1 °C to 89.7 °C. Blanc et al. (2013) collect a set of solubility experimental data related to kaolinite previously presented by various authors and associated with temperature values ranging between 25 °C and 300 °C.

We employ the following procedure to quantify uncertainties associated with $\tilde{A}_k$ and $\tilde{D}_k$. (with $i$ = kaolinite, calcite) using the solubility experimental data indicated above:

1. We calibrate model (5) against available experimental observations $K_{Sk}^*$ upon estimating the parameters $\tilde{A}_k$ and $\tilde{D}_k$ (with $k$ = calcite, kaolinite) through a standard least square criterion.

As indicated above, parameters $B_k$, $C_k$, $E_k$ are set to the corresponding values reported in Thermoddem. This procedure yields best estimates ($\hat{\tilde{A}}_k$, $\hat{\tilde{D}}_k$) of parameters ($\tilde{A}_k$, $\tilde{D}_k$) and the related uncertainty expressed in terms of a 2×2 symmetric covariance matrix $\hat{\boldsymbol{\Psi}}_k$. The results of these calculations are listed in the first two rows of Table 5.

2. We consider that the entries of the uncertain parameter vectors $\tilde{\boldsymbol{u}}_k = (\tilde{A}_k, \tilde{D}_k)$ can be described through a bivariate Gaussian distribution with mean $\mu(\tilde{\boldsymbol{u}}) = (\hat{\tilde{A}}_k, \hat{\tilde{D}}_k)$ and covariance matrix $\boldsymbol{\Psi}_k = \hat{\boldsymbol{\Psi}}_k$ (with $k$ = calcite, kaolinite).

No direct references are reported in the Thermoddem database to characterize the uncertainty associated with the equilibrium constants related to the remaining phases included in (8) (dolomite, clinochlore, quartz, $CO_{2(g)}$, $CO_{2(aq)}$). In our illustrative example we resort to the following set of assumptions to characterize uncertainties associated with $\tilde{A}_k$ and $\tilde{D}_k$ (with $k$ = dolomite, clinochlore, quartz, $CO_{2(g)}$, $CO_{2(aq)}$):

1. The vector of parameters $\tilde{\boldsymbol{u}}_k = (\tilde{A}_k, \tilde{D}_k)$ (with $k$ = dolomite, clinochlore, quartz, $CO_{2(g)}$, $CO_{2(aq)}$) is associated with a bivariate Gaussian distribution. Here, we assume that the entries of the vector of mean values $\mu(\tilde{\boldsymbol{u}}_k)$ coincide with the values included in Thermoddem for each phase $k$ (see Table 5).

2. Affine minerals are characterized by the same parametric uncertainty, i.e., we set $\boldsymbol{\Psi}_{dolomite} = \hat{\boldsymbol{\Psi}}_{calcite}$ (as dolomite and calcite are both carbonates minerals) and $\boldsymbol{\Psi}_{clinochlore} = \hat{\boldsymbol{\Psi}}_{kaolinite}$ (as clinochlore and kaolinite are both clay minerals, Bergaya and Lagaly, 2013).

3. The parameters describing the solubility of quartz and the water transition phase equilibrium are characterized by negligible uncertainty when compared against the uncertainty level of the equilibrium constants discussed above.

4. We set $\Psi_{CO2(g)} = \Psi_{CO2(aq)} = \hat{\Psi}_{kaolinite}$, as $\hat{\Psi}_{kaolinite}$ renders the highest level of uncertainty following estimation of the coefficients of (5) through the experimental data employed (i.e., solubility data of calcite and kaolinite).

Mean values of the parameter distributions are listed in Table 5 together with the associated covariance matrix entries and the set of assumptions illustrated above. Note that these assumptions are not strictly required for the applicability of the proposed methodology and are here considered solely for illustrative purposes. In this sense, measurements on mineral solubility or equilibrium constants can readily be integrated in the proposed workflow when available.

## 2.3 Quantitative assessment of $CO_2$ generation and CCR mechanism activation at basin scale

The basin compaction and geochemical models illustrated in Sections 2.1 and 2.2 allow assessing the desired dynamics of the CCR mechanism and quantifying the amount of $CO_2$ generated (as gaseous or dissolved species) during the diagenetic process.

Generation of a separate gas phase at a location $Z$ and time $t$ takes place when

$$\tilde{R}(Z,t) = \frac{\tilde{P}_{gas}(Z,t)}{P(Z,t)} = \frac{\tilde{P}_{CO2}(Z,t) + P_{H2O}(Z,t)}{P(Z,t)} \geq 1 \qquad (11)$$

where $P(Z,t)$ and $\tilde{P}_{gas}(Z,t)$ respectively are the fluid pressure and the partial pressure of the gas phase. Note that $P(Z,t)$ is rendered by the basin compaction model of choice (see Section 2.1) while $\tilde{P}_{CO2}$ is evaluated through the procedure illustrated in Section 2.2. The computation of partial pressure of water vapor, $P_{H2O}(Z,t)$, is detailed in Electronic Annex I. According to criterion (11), the space-time locations at which the generation of gaseous $CO_2$ may take place can be identified through the local values of the ratio $\tilde{R}$. For a given time level $t$, the activation of the mechanism is assigned to the location

$$\tilde{Z}_{act}(t) = \begin{cases} \varnothing & \text{if } \tilde{R} < 1 \text{ for all } Z \in \Omega_Z \\ \min\{Z \in \Omega_Z | \tilde{R}(Z,t) \geq 1, m_{CO2} > 0\} & \text{if } \exists Z \in \Omega_Z | \tilde{R}(Z,t) \geq 1 \end{cases} \quad (12)$$

i.e., the location of the $CO_2$ source at time $t$, $\tilde{Z}_{act}(t)$, is assumed to correspond to the shallowest depth at which $\tilde{R} \geq 1$, given that the mineral composition is compatible with CCR. Note that $\tilde{Z}_{act}(t)$ is a function of time because of the temporal variability of vertical profiles of temperature and pressure. Definition (12) is consistent with the assumption that $CO_2$ migrates instantaneously upwards when a gas phase is formed (see Fig. 1). Under such conditions, the primary phases of the equilibrium reaction (1) are progressively consumed because one of the secondary phases ($CO_2$) is continuously driven away. This behavior is observed until the limiting reactant in (1) vanishes. We assume that the complete consumption of at least one primary phase takes place on a time scale that is considerably smaller than the one associated with the basin evolution. Therefore, the burial velocity of the sediment, $V_{SED}(\tilde{Z}_{act}, t)$, is a limiting factor for the generation of $CO_2$ as a gas phase through a CCR mechanism. Under this assumption, we can then evaluate the rate of $CO_2$ generation as

$$\tilde{F}_{CO2}(t) = m_{CO2} \cdot V_{SED}(\tilde{Z}_{act}, t) \cdot [1 - \phi(\tilde{Z}_{act}, t)] \cdot L \cdot \rho \quad (13)$$

Here, $\tilde{F}_{CO2}$ [kg/Ma] is the $CO_2$ mass generation rate; $L$ [m$^2$] is the planar cross sectional area of the basin/reservoir; and $m_{CO2}$ [-] is maximum amount of mass of $CO_2$ released by unity mass of sediment (see Electronic Annex I for further detail about the computation of $m_{CO2}$). Note that, following (8), the limiting reactant is dolomite in the two mineralogical scenarios investigated in this work (Table 3). When the gas generation mechanism is activated, the reaction evolves over time until at least one primary phase is exhausted (see Section 1). We note that $\tilde{F}_{CO2} \equiv 0$ when the mechanism is not activated (i.e., when $\tilde{Z}_{act}(t) = \varnothing$).

According to the conceptual model described above, at least one of the mineral phases involved in the CCR mechanism is expected to be exhausted at locations below the activation depth (i.e, for $Z > \tilde{Z}_{act}$) and the mineral phases equilibrium (1) leading to dissolved $CO_2$ is no longer possible. We therefore assume that the dissolved amount of $CO_2$ is zero at all locations $Z > \tilde{Z}_{act}$ (see Electronic Annex I for additional details about the computational procedure).

## 3 Analysis of sources of uncertainty

Any model which aims at quantifying $CO_2$ generation in sedimentary basin is subject to considerable uncertainties. These are due to our incomplete knowledge of the processes involved and of the initial/boundary conditions together with the lack of information resulting from the large space-time scales, which are characteristic of the evolution of sedimentary systems. Upon following Neuman (2003), we distinguish in the following sections between modeling and parametric uncertainties. This work is keyed to the development and implementation of a methodology for the quantification of the uncertainty stemming from our incomplete knowledge of equilibrium reaction constants. In this section we frame this choice within the context of uncertainty quantification and discuss a variety of possible sources of uncertainty which may be relevant to our setting.

### 3.1 Model uncertainties

Investigation of complex settings in earth and environmental sciences typically relies on the formulation of a conceptual-mathematical model which is consistent with available information on the system investigated. Multiple and competing conceptual models can be formulated, according to diverse interpretations of the processes underlying the target scenario.

We list here key model uncertainties and the related assumptions associated with our setting.

- While we focus on the occurrence of reaction (8), other geochemical processes may take place simultaneously during basin compaction. Different competing models could therefore be formulated according to which reaction (8) occurs jointly with a set of diagenetic processes (e.g., dolomitization, albitization, illitization, cracking of biological matter and many others). All these processes can jointly contribute to $CO_2$ partial pressure and to increase/decrease or to the amount of $CO_2$ which can be found in the system. The selection of the geochemical processes which should be considered and the formulation of a related model is not a trivial task and constitutes a remarkable source of model uncertainty.

- The selection of the primary phases considered in the mineralogical assemblage is a key input to our methodology. This information is typically uncertain and various admissible hypotheses may be formulated, consistent with geological and sedimentological conceptual models and interpretations. Companion considerations hold on the assumed initial interstitial fluids composition (i.e., gas phase and brine). Our approach rests on the assumption that *i*) gaseous phase are $CO_2$ and $H_2O$, and *ii*) the initial pore-brine is pure water and the primary phases appearing in (8) are all available in the mineral composition. This is a simplification of the conditions encountered in real cases, but does not disable the proposed methodology.

- The spatial arrangement of the mineral composition may be affected by heterogeneity at all scales. In our conceptual model we assume a uniform spatial distribution of primary mineral phases throughout the carbonate-rich sedimentary layers. The spatial/temporal distribution of minerals could alternatively be described as a stochastic process, whose main features should possibly be characterized through real mineralogical samples of a specific sedimentary basin case study.

Quantification of the modeling uncertainties listed above may be performed through dedicated techniques (see, e.g., Neuman, 2003). While this task lies beyond the scope of the present work, we remark that these types of uncertainties should be carefully considered prior to applying the procedure outlined in Section 2 to the interpretation of observations from a real field site.

## 3.2 Parametric uncertainties

An admissible conceptual/mathematical model of a process commonly includes a number of parameters. These are in turn associated with a given level of uncertainty due to lack of information. This incomplete knowledge about parameter values can be quantified through, e.g. statistical characterization of available experimental data via parameter estimation techniques. In this work, we *a*) present a rigorous methodology to account for parametric uncertainty associated with mineral solubility equilibrium constants and *b*) propagate such uncertainty throughout our geochemical model of choice, which is aimed at representing CCR. While the need to account for these parameters is ubiquitous in geochemical models of environmental systems, a rigorous quantification of their uncertainty and its ensuing effects is often neglected. To sharply delineate the effect of this specific source of parametric uncertainty, we do not consider here other sources of parametric uncertainties such as: *a*) properties of the sedimentary rocks, i.e., density, permeability, thermal diffusivity and mechanical compressibility; *b*) boundary conditions of the compaction problem, i.e., heat flux at basin basement, and temporal dynamics of sea level evolution and sediment deposition rate; and *c*) other parameters of the geochemical model, including relative abundance of each mineral phase in the primary assemblage (for the given the qualitative composition of the mineralogy, which falls into the category of modeling uncertainties, as discussed in Section 3.1), molar volume change during the reaction (8), and activity and fugacity coefficients.

The influence of each set of parameters may be assessed through local and/or global sensitivity analysis techniques (e.g., Razavi and Gupta, 2015 and references therein), which we envision to explore in future works.

## 4 Results

This Section is devoted to a synthetic illustration of the results stemming from the implementation of the methodology proposed in Section 2.

## 4.1 Basin evolution

The results depicted in Fig. 3 are obtained through the numerical solution of the basin compaction model illustrated in Section 2.1. Fig. 3a depicts the space-time evolution of the vertical stratigraphic sequence of the basin (i.e., the system geo-history). The total basin thickness at present day is also shown. Fig. 3b-d respectively depict the space-time evolution of porosity, temperature and pressure with reference to the stratigraphy displayed in Fig. 3a. The black vertical lines identify the times when the sediment deposition rate ($V_D$) changes its value according to stepwise function described in Table 2 imposed at basin top.

## 4.2 Consistency of geochemical modeling results with field data

Here, we compare the results stemming from the application of our geochemical modeling approach against a set of field observations of $CO_2$ partial pressures reported by Coudrain-Ribstein et al. (1998). This comparison aims at assessing the robustness of our procedure and of the assumptions underlying the uncertainty quantification steps proposed in Section 2.2.1. We focus on the variation of $\log \tilde{K}_{R,T,P}$ and $\log \tilde{P}_{CO2}$ as a function of temperature and pressure.

To this end, we perform a Monte Carlo sampling of the parameter space to obtain $N$ realizations (here, we consider $N = 10^5$) of $\log \tilde{K}_{R,T,P}$ (6) as function of temperature and pressure. In the context of our comparison between field data and geochemical model outputs, we assume the following relationship between temperature and pressure (Smith and Ehrenberg, 1989; Cathles and Schoell, 2007)

$$P[bar] = 6(T[K] - 298) \tag{14}$$

Fig. 4a depicts the dependence on temperature of the mean, median, and 1$^{st}$- and 99$^{th}$-percentiles of the sample distribution of $\log \tilde{K}_{R,T,P}$. Here and in the following we denote a percentile (or quantile) of the distribution of a random variable $\tilde{\zeta}$ as $p_W(\tilde{\zeta})$. The latter is defined as the value below which a percentage equal to $W$ of observations of $\tilde{\varsigma}$ falls. Note that the mean and the median coincide in Fig. 4a, $\log \tilde{K}_{R,T,P}$ being characterized by a symmetric sample distribution.

The Monte Carlo sample of $\log \tilde{P}_{CO2}$ values can be obtained from $\log \tilde{K}_{R,T,P}$ through (3). Fig. 4b depicts percentiles $p_1(\log \tilde{P}_{CO2})$, $p_{50}(\log \tilde{P}_{CO2})$, and $p_{99}(\log \tilde{P}_{CO2})$ as a function of temperature. These Monte Carlo - based results are juxtaposed in Fig. 4b to a set of available measurements of $P_{CO2}$ reported by Coudrain-Ribstein et al. (1998) for sedimentary formations. The consistency of the results provided by our geochemical model and the field data depicted in Fig. 4b is discussed in Section 5.1.

## 4.3 Quantitative assessment of $CO_2$ generation and CCR mechanism activation

We present here results associated with the way parametric uncertainty propagates to the outputs of the model described in Section 2.3, i.e., to the rate of generation of gaseous $CO_2$ and to the total dissolved $CO_2$. Characterization of parameter uncertainty relies on the procedure described in Section 2.2.1. The results are related to the mineral compositions $S_{dol}$ and $S_{cal}$ (see Table 4) and are discussed in Section 5.2. All of the results presented are obtained upon relying on a sample of $N = 10^5$ Monte Carlo realizations.

Fig. 5 shows the vertical profiles of the percentiles of the partial pressure of $CO_2$, $p_W(\log \tilde{P}_{CO2} | C_{max})$, and of the ratio $\tilde{R}$ as defined in (11), $p_W(\tilde{R})$ ($W$ = 1, 25, 50, 75, 99), at two selected time levels ($t$ = 48, 0 Ma) and for scenarios $S_{dol}$ and $S_{cal}$. To complement this result, Fig. 6 provides a comparison of the sample *cdf* (cumulative distribution function) of $\tilde{R}$ (11) at $Z$ = 8 km for scenarios $S_{dol}$ and $S_{cal}$. Note that at $t$ = 0 the top layer (0 < $Z$ < 1.4 km) of the basin is formed by mudrocks (see Fig. 3a). Therefore, we set $P_{CO2} = 0$ at these locations (see Fig. 5b and d), as we assume CCR happens exclusively in carbonates layers.

The probability of activation $G_A(t)$ can then be as the sample probability (relative frequency) of observing at least one point in the domain for which $\tilde{R} \geq 1$, i.e., the generation of $CO_2$ as a separate phase through CCR is activated at time $t$. The procedure to compute GA is exemplified in Fig. 6, where the value $R = 1$ is identified by a vertical red line, which represents the conditions at which the $CO_2$ generation as a separate gas phase is activated (see Section 2.3). The *cdfs* associated with the two diverse mineral compositions intercept the threshold line corresponding to $R = 1$ (i.e., the conditions at which the $CO_2$ generation as a separate gas phase is activated) at different points, i.e., for $R = 1$ the *cdf* attains a value equal to 0.55 and 0.85, respectively for $S_{dol}$ and $S_{cal}$, indicating a different probability of activation in the two scenarios. Fig. 7 depicts the temporal evolution of $G_A(t)$ for $S_{dol}$ and $S_{cal}$ across the overall basin history.

Our procedure allows identifying not only the probability of activation at given time but also to estimate the location of the $CO_2$ sources through (12). Fig. 8 depicts the sample probability (relative frequency) $f_{Zact,t}$ that the activation of gaseous $CO_2$ generation takes place at location $\tilde{Z}_{act}$ at time $t$. In particular Fig. 8a displays $f_{Zact,10}$, i.e., $f_{Zact,t}$ for $t = 10$ Ma, where the domain is comprised between the sea bottom (at 106 m) and 7.6 Km. We note that this relative frequency is computed upon considering the complete set of Monte Carlo realizations, including those for which $\tilde{Z}_{act} = \emptyset$ according to (12). Thus, the function $f_{Zact,t}$ integrates to the corresponding value of $G_A$ at time $t$ i.e.

$$\int_{\Omega_Z(t)} f_{Zact,t} dZ = G_A(t) \tag{15}$$

For example, the integral (15) evaluated at $t = 10$ Ma for scenario $S_{dol}$ is equal to 0.30, which corresponds to the value of $G_A(t=10$ Ma$)$ for the corresponding scenario reported in Fig. 7. Fig. 8b-c depict the temporal dynamics of the relative frequency $f_{Zact,t}$ for scenarios $S_{dol}$ and $S_{cal}$. As anticipated by the temporal variation of $G_A(t)$ in Fig. 7 the nonzero values are obtained for $t < 50$ Ma in both scenarios. The generation of gaseous $CO_2$ takes place at $\tilde{Z}_{act} > 4.8$ km in the considered example for both mineral composition scenarios.

Figure 9-10 provide the probabilistic quantification of the generated $CO_2$ in terms of *(i)* flux of gaseous $CO_2$ generated as a result of the CCR process. $\tilde{F}_{CO2}$, as defined in (13), and *(ii)* concentration of dissolved $CO_2$ $\tilde{C}_{CO2(aq)}|(C_{max},\tilde{Z}_{act})$. Fig. 9a depicts the relative frequency $f_{F,t}$ associated with $\log \tilde{F}_{CO2}$ at time $t$ for the overall basin history of scenario $S_{dol}$. Corresponding results for $S_{cal}$ are depicted in Fig. 9b. We set here $L = 1$ m$^2$ in (13) so that the reported values of $\tilde{F}_{CO2}$ are per unit (planar) area of the sedimentary basin. For completeness, Fig. 9a-b include the information (black solid curve) corresponding to the frequency of activation $G_A(t)$. Note that $G_A(t) \equiv 0$ for $t \in ]\,45 \text{ Ma}, 135 \text{ Ma}]$, thus implying that $\tilde{F}_{CO2} \equiv 0$ across all Monte Carlo realizations for these simulation times. Indeed, $f_{F,t} = 0$ for all non-zero values of $\tilde{F}_{CO2}$ for $t \in ]\,45 \text{ Ma}, 135 \text{ Ma}]$. Fig. 9c depicts the sample *cdf*s of $\tilde{F}_{CO2}$ associated with the two time levels identified by the red dashed vertical lines in Fig. 9a, i.e., $t = 20$, and 0 Ma.

Figure 10 reports the distribution along the basin depth of the relative frequency associated with the log-concentration $\log \tilde{C}_{CO2(aq)}|(C_{max},\tilde{Z}_{act})$ (denoted as $f_{C,Z}$ in Fig. 10a-b). Introducing here $f_{C,Z}(0)$ to denote the relative frequency associated with $\tilde{C}_{CO2(aq)}|(C_{max},\tilde{Z}_{act}) = 0$, Fig. 10c, d) display the variation of $f_{C,Z}(0)$ with $Z$ for the two mineral compositions $S_{dol}$ and $S_{cal}$, respectively.

## 5 Discussion

This Section is devoted to the discussion and interpretation of the results illustrated in Section 4. We focus in particular on the key results obtained in terms of the probabilistic assessment of $CO_2$ generation through CCR.

## 5.1 Geochemical modeling results

With reference to Fig. 4a, we observe that all percentiles associated with $\log \tilde{K}_{R,T,P}$ tend to increase with temperature and pressure. Fig. 4a shows that a negligible probability is associated with positive values of $\log \tilde{K}_{R,T,P}$ when $T < 50$ °C, i.e. the equilibrium (8) favors primary phases over secondary phases. Otherwise, our results indicate that a probability very close to 1 is associated with values of $\log \tilde{K}_{R,T,P} > 0$ for $T > 100$ °C. This finding is consistent with the results of Smith and Ehrenberg (1989) who suggest that $CO_2$ formation is typically favored above 100-120°C as a consequence of carbonate phase consumption.

Fig. 4b, shows that the partial pressure of $CO_2$ tends to increase with temperature as a direct consequence of the trend of $\log \tilde{K}_{R,T,P}$ in Fig. 4a. The median value of $\log \tilde{P}_{CO2}$, $p_{50}\left(\log \tilde{P}_{CO2}\right)$ is consistent with field observations (e.g., Texas, Norway and Thailand basins in Fig. 4b) for temperature values higher than 100 °C. Almost all of the field data reported by Coudrain-Ribstein et al. (1998) for this temperature range fall between $p_1\left(\log \tilde{P}_{CO2}\right)$ and $p_{99}\left(\log \tilde{P}_{CO2}\right)$, with the exception of a very limited number of points. Otherwise, the majority of the field data (mainly associated with Alberta, Paris, Arkansans and Medison basins in Fig. 4b) falls outside the range identified by $p_1\left(\log \tilde{P}_{CO2}\right)$ and $p_{99}\left(\log \tilde{P}_{CO2}\right)$ for $T < 100$ °C. The median value of $\log \tilde{P}_{CO2}$ resulting from our simulations tends to overestimate the field data in this temperature range. Giggenbach (1981) suggests that dilution of aqueous $CO_2$ in the system at shallow depth (corresponding to low temperature) can happen due to mixing of fresh and cold water (i.e., from meteoric precipitations) with groundwater. Moreover, Coudrain- Ribstein et al. (1998) observe that complex minerals such as illite or competing geochemical processes can play a relevant role at low temperature levels. The discussion of the consistency of the data with possible alternatives of physical and conceptual models as the ones suggested above is beyond the scope of the present work as previously explained in Section 3.1. Here, we can highlight that our procedure leads to results which are consistent with the degree of variability of $P_{CO2}$ values observed in real systems at temperatures $T > 100$ °C.

## 5.2 $CO_2$ generation and CCR mechanism activation

We start our discussion by considering the characterization of $\tilde{P}_{CO2}$ as function of depth. All values of $p_W\left(\log \tilde{P}_{CO2} | C_{max}\right)$ display a monotonic increase with depth (Fig. 5a-b) at the considered times and for both mineralogical composition scenarios. This behavior is consistent with the observation that (*i*) temperature and pressure increase with depth at all times (see Fig. 3c and d); and (*ii*) the equilibrium constant $\log \tilde{K}_{R,T,P}$ increases with temperature and pressure (see Fig. 4a), i.e., formation of $CO_2$ is favored by the increase of temperature and pressure. Partial pressure of $CO_2$ is computed only in those layers within which there is a mineral composition compatible with the CCR process, labeled as carbonate layers in Figure 3. The total basin thickness at $t = 48$ Ma is approximately equal to 5.5 km, the basin being completely constituted by carbonates rocks (see Fig. 4a). Thus, we find $\tilde{P}_{CO2} | C_{max} > 0$ across the whole computational domain (Fig. 5a). The impact of the limiting reactant associated with the two mineral composition scenarios is negligible at this time level and no significant differences are detected between values of $p_W\left(\log \tilde{P}_{CO2} | C_{max}\right)$ computed for scenarios $S_{cal}$ and $S_{dol}$. We can then conclude that the dissolved $CO_2$ concentration values rendered by the geochemical model at this time do not exceed the value of the maximum admissible concentration associated with either $S_{dol}$ or $S_{cal}$. Otherwise, the mineral composition at $t = 0$ Ma influences the statistical characterization of $\log \tilde{P}_{CO2} | C_{max}$ at large depths ($Z > 6$ km). We observe that $p_W\left(\log \tilde{P}_{CO2} | C_{max}\right)$ displays a different trend for depths larger than 6 km, according to the mineralogical composition considered. Fig. 6b suggests that the effect of limiting reactant affects all probability levels, i.e., $p_W\left(\log \tilde{P}_{CO2} | C_{max}(S_{dol})\right) > p_W\left(\log \tilde{P}_{CO2} | C_{max}(S_{cal})\right)$ for all considered values of $W$ even as the value of $C_{max}$ (quantifying the effect of limiting reactant) is a deterministically imposed upper boundary (see the Electronic Annex III for additional details).

We then discuss the results obtained in terms of the activation of the generation of gaseous $CO_2$. Values of $\tilde{R}$ associated with all of the considered percentiles $p_I(\tilde{R})$ increase with depth for both time levels considered (see Fig. 5c-d). This result indicates that the sum of gas partial pressures ($\tilde{P}_{CO2}$ and $P_{H2O}$) tends to increase with depth at a faster rate than does the fluid pressure $P$. The difference $p_{99}(\tilde{R}) - p_1(\tilde{R})$ markedly increases with depth, suggesting that the level of uncertainty associated with $\tilde{P}_{CO2}$ tends to increase with temperature and pressure. Consistent with Fig. 5b, the mineral composition scenario influences these results only for $Z > 6$ km. Fig. 6 presents the comparison of the sample *cdfs* (cumulative distribution functions) of $\tilde{R}$ (11) at $Z = 8$ km for scenarios $S_{dol}$ and $S_{cal}$. We observe that the relative proportions among the different minerals constituting the sediments influences the statistical distribution of $\tilde{R}$ and, consequently, the probability of generation of gaseous $CO_2$.

The value of the sample probability of generation of gaseous $CO_2$ $G_A(t)$ increases with time (Fig. 7) and attains its highest value for the final simulation time ($t = 0$ Ma). It is possible to distinguish three stages according to the time evolution of $G_A(t)$: (*i*) for $t \in\ ]45\text{ Ma}, 135\text{ Ma}]$, where $G \equiv 0$; *ii*) for $t \in\ ]20\text{ Ma}, 45\text{ Ma}]$, where $0 < G < 0.2$, with comparable values for $S_{dol}$ and $S_{cal}$; and (*iii*) for $t \in\ ]0\text{ Ma}, 20\text{ Ma}]$, where $G_A$ continuously increases, with a trend which varies according to the mineralogical scenario. It can be noted that the probability of activation grows slower in time for scenario $S_{cal}$ than for $S_{dol}$.

Our results suggest that the temperature range associated with locations where the activation of the process is possible, i.e., at which $f_{Zact,\tau} > 0$, is comprised between 200 and 300 °C (compare Fig. 8b-c with Fig. 4c). This information can be highly valuable, e.g., to assess the prior probability of CCR being a key source of $CO_2$ in natural systems (e.g., Jarvie and Jarvie, 2007). Cathles and Schoell (2007) predict an activation temperature of 330°C through a deterministic approach similar to the one presented in Section 2 and a simple time-independent *P-T* relationship. Our results suggest that the generation of gaseous $CO_2$ by CCR might take place also at lower temperatures when the parametric uncertainty related to the geochemical model are considered.

With reference to the results depicted in Fig. 9 and related to the probabilistic analysis of $\tilde{F}_{CO2}$, we note that nonzero (positive) values of the latter can be found only if the CCR mechanism is active at a given time, i.e., if $\tilde{Z}_{act}(t) \neq \varnothing$, $\tilde{F}_{CO2}$ being equal to zero otherwise. The contour lines describing $f_{F,t}$ in Fig. 9a-b are qualitatively very similar. However, we observe a remarkable quantitative difference between the two scenarios analyzed: non-zero values of $\tilde{F}_{CO2}$ range between 12 and 31 ton/Ma in scenario $S_{dol}$, while these are comprised between 1.0 and 2.5 ton/Ma for $S_{cal}$. This result can be ascribed to the effect of the diverse fractions of dolomite characterizing $S_{dol}$ and $S_{cal}$ and acting as the limiting reactant. The *cdfs* reported in Fig. 9c indicate that the nonzero values of $\tilde{F}_{CO2}$ display a modest variability for a given time level. This suggests that, even as the location of the source is characterized by remarkable variability across the Monte Carlo sample (see Fig. 8), porosity and sediment velocity which contribute to $\tilde{F}_{CO2}$ according to (13) display modest variability along the region of vertical domain where $f_{Zact,t} > 0$. Our results also show that the nonzero values of $\tilde{F}_{CO2}$ observed at *t* = 20 Ma in the Monte Carlo sample are fewer than those obtained at *t* = 0 Ma. Note that the non-zero values of $\tilde{F}_{CO2}$ detected at *t* = 20 Ma are larger than their non-zero counterparts arising at *t*=0 Ma. For those realizations within which the generation of gaseous $CO_2$ is activated, we obtain a $CO_2$ generation rate of about 27 and 18 ton/Ma, respectively at *t* = 20, and 0 Ma. This difference is a consequence of the diverse values of the sediment burial velocity ($V_{SED}(\tilde{Z}_{act}, t)$ in (13)) at the location where gaseous $CO_2$ is generated. We exclude that porosity can play a relevant role in the different $\tilde{F}_{CO2}$ values obtained at *t* = 0 and 20 Ma as it is almost constant (approximately equal to 0.1) for *Z* > 4 km, where the $CO_2$ source is located (see Fig. 3c).

We conclude our discussion by considering the distribution of $\log \tilde{C}_{CO2(aq)} | (C_{max}, \tilde{Z}_{act})$ (Eq.I.23 in the Electronic Annex I) depicted in Fig. 10. We recall that quantity $f_{C,Z}(0)$ denotes the relative frequency associated with $\tilde{C}_{CO2(aq)} | (C_{max}, \tilde{Z}_{act}) = 0$. Fig. 10c, d) respectively depict the dependence of $f_{C,Z}(0)$ on $Z$ for $S_{dol}$ and $S_{cal}$. Note that, according to our conceptual model, $\tilde{C}_{CO2(aq)} | (C_{max}, \tilde{Z}_{act}) = 0$ at all locations where mudstone layers are found and below the depth $Z_{act}$. As such, we find $f_{C,Z}(0) = 1$ at $0 < Z < 1.4$ km. We observe that $f_{C,Z}(0) \equiv 0$ at $1.4 < Z < 4.5$ km, suggesting that the concentration of dissolved $CO_2$ attains non-zero values across the complete Monte Carlo set. Finally, the relative frequency $f_{C,Z}(0)$ attains values higher than zero and lower than one and increases with depth for $Z > 4.5$ km. This finding is consistent with results of Fig. 8, showing that (*i*) $Z = 4.5$ km is the shallowest location at which the activation of the CCR mechanism is possible at $t = 0$ Ma and (*ii*) the probability to observe vanishing $CO_2$ concentrations at a given location increases with the relative frequency that the depth of such a location is larger than that corresponding to $Z_{act}$.

Calculated values for concentration of dissolved $CO_2$ display negligible dependence on mineral composition scenario, in contrast with $\tilde{F}_{CO2}$ (Fig. 9). The only impact of the mineral composition scenario on $f_{C,Z}$ is due to the upper bound $C_{max}$ imposed by the availability of reactants which leads to an increase of the relative frequency $f_{C,Z}$ of values $\tilde{C}_{CO2(aq)} | (C_{max}, \tilde{Z}_{act}) = C_{max}$ at large depths within $S_{cal}$ (see Fig. 10b). This behavior follows from the observation that the extent of the region where the reaction can occur is limited by the available dolomite volume fraction in $S_{cal}$ (see Fig. 10d).

# 6 Conclusions

We present a methodology conducive to a probabilistic assessment of the amount of $CO_2$ generated in sedimentary basins as consequence of the interaction between carbonate and clay minerals in the presence of pore-water. Our modeling strategy rests on the quantification of the uncertainty of chemical equilibrium parameters related to mineral solubility and the way it propagates to key model outputs. Application of the proposed workflow leads to a probabilistic assessment of: (*i*) the evolution of $CO_2$ partial pressure and dissolved $CO_2$ as a function of depth and time along the basin burial history; (*ii*) the location of the source where gaseous $CO_2$ is released from the sediments; (*iii*) the amount of gaseous $CO_2$ released per unit time.

We illustrate our approach upon relying on a realistic basin compaction history meaning that temperature-pressure-porosity combinations are compatible with realistic fields. Our work provides a first attempt to quantify $CO_2$ generation by CCR at geological scales with the explicit inclusion of a probabilistic assessment of the uncertainty stemming from the incomplete knowledge of mineral solubility and phase equilibrium constants at high temperatures. Due to its flexibility, we envision that the framework proposed here can be readily extended to include the uncertainty related to the basin pressure and temperature dynamics. We envision that the proposed model may be extended in future works to include other sources of model uncertainty, such as those associated with pore-water chemistry (e.g., salinity).

Our uncertainty quantification is based on data of mineral solubility and phase equilibrium constants available at laboratory scales. We verify that the procedure we employ to characterize parametric uncertainty of the geochemical model leads to results which are consistent with field observations of $CO_2$ partial pressure in sedimentary formations reported in the literature.

Our study shows that the partial pressure of $CO_2$ displays a monotonic and increasing trend with depth. This suggests that the increase of temperature taking place during a basin burial history favors the progressive generation of $CO_2$ at the expense of carbonate mineral phases. $CO_2$ is generated as a separate phase only under specific conditions which depend on temperature and pressure distributions. The probability that these conditions are encountered tends to increase with time and attains its largest value (around 0.45 in the setting we analyze) at the end of the simulation period, which represent the present day. In our example we find that generation of $CO_2$ through CCR can become effective at temperatures comprised between 200 and 300 °C. These specific results are conditional to the given compaction history of the basin and of the geochemical model structure selected in this study are therefore not amenable to direct transferability to diverse geological settings.

Mineral compositions associated with sediments largely affect the flux of generated $CO_2$. In the case we examine, the key driver is the amount of dolomite associated with the sediments and representing the source of $CO_2$. In our illustrative example, the impact of model parameter uncertainty is stronger on the activation depth than on the $CO_2$ generation rate. As a consequence, our findings suggest that reliable estimates of $CO_2$ migration scenarios should rely on accurate characterization of mineral composition as well as geochemical model parameters.

# 7   Acknowledgments

We acknowledge financial support by Eni spa.